\documentstyle[aps,prl,epsfig]{revtex}

\begin{document}

\twocolumn[\hsize\textwidth\columnwidth\hsize\csname
@twocolumnfalse\endcsname
\title{Phase Diagrams from Topological Transitions: \\
The Hubbard Chain with Correlated Hopping}
\author{A. A. Aligia$^{a}$, K. Hallberg$^{a}$, C. D. Batista$^{a}$, and G.
 Ortiz$^{b}$}
\address{$^{a}$Comisi\'{o}n Nacional de Energ{\'{\i }}a At\'{o}mica, 
Centro At\'omico Bariloche and Instituto Balseiro, \\
8400 S.C. de Bariloche, Argentina}
\address{$^{b}$Theoretical Division, 
Los Alamos National Laboratory, Los Alamos, NM 87545}

\date{Received \today }

\maketitle

\begin{abstract}
The quantum phase diagram of the Hubbard chain with  
correlated hopping is accurately determined through jumps in $\pi$ in the 
charge and spin Berry phases. The nature of each thermodynamic phase, and the 
existence of charge and spin gaps, is confirmed by calculating correlation 
functions and other fundamental quantities using numerical methods, and 
symmetry arguments. Remarkably we find  
striking similarities between the stable phases for moderate on-site Coulomb 
repulsion: spin Peierls, spin-density-wave and triplet superconductor, and 
those measured in (TMTSF)$_2$X.
\end{abstract}  

\pacs{Pacs Numbers: 71.10.+x, 3.65.Bz, 71.27.+a}

\vspace*{-1.4cm}

\vskip2pc]

\narrowtext

The search for electronic mechanisms of superconductivity and the study of 
superconducting and Mott phase transitions are among the most interesting 
subjects of the physics of strongly correlated systems. In few cases, exact 
results have helped to elucidate the nature of these transitions 
\cite{sha,str,afg}. In general one has to rely on numerical 
calculations of finite systems for which quantities like the Drude weight 
$D_c$ (which should vanish for an insulator in the thermodynamic limit 
\cite{kohn}), or any other correlation function, vary smoothly at the 
transition. Consequently, for instance, the boundaries between 
a charge-density-wave (CDW) or spin-density-wave (SDW) insulators and metallic 
phases in half-filled generalized Hubbard models were difficult to establish 
\cite{hir,mit}. 

The Berry phase is a general geometrical concept which finds 
realizations in various physical problems \cite{berry}. 
It is the anholonomy associated to the parallel transport of 
a vector state in a certain parameter space. In condensed matter, the 
charge Berry phase $\gamma_c$ is a measure of the macroscopic electric 
polarization in band or Mott insulators \cite{ort} while the spin Berry phase 
$\gamma_s$ represents its spin polarization \cite{gag,epl}. In systems with inversion 
symmetry $\gamma_c$ and $\gamma_s$ can attain only two values: 0 or $\pi$ 
(mod$(2 \pi)$). Thus, if two thermodynamic phases differ in the topological 
vector $\vec{\gamma} = (\gamma_c,\gamma_s)$ this sharp difference 
allows us to unambiguously identify the transition point even in finite 
systems. This ``order parameter'' was recently used to detect 
metallic, insulator and metal-insulator transitions in one-dimensional 
lattice fermion models \cite{gag,epl}. 

In this Letter we determine the quantum phase diagram of the Hubbard chain 
with correlated hopping at half-filling using topological transitions. The 
phase diagram is very rich showing two metallic and two insulating 
thermodynamic phases each characterized by one of the four possible values of 
the topological vector $\vec{\gamma}$. One of the metallic phases corresponds 
to a Tomonaga-Luttinger liquid with dominant triplet superconducting  
correlations at large distances (TS). This is interesting since 
there is experimental evidence 
indicating that the Bechgaard salts (TMTSF)$_2$ClO$_4$ and (TMTSF)PF$_6$ 
under pressure are TS  \cite{tak,lee,bel}. Furthermore, the 
insulating SDW and spin gapped spin-Peierls phase observed in (TMTSF)PF$_6$
as the pressure is lowered \cite{jer} are also present in the model phase 
diagram.   

The effective model Hamiltonian is: 
\cite{afg}

\begin{eqnarray}
&H& = \sum_{\langle i,j \rangle \sigma} ( c^{\dagger}_{i \sigma} c^{\;}_{j 
\sigma} + h.c. ) \left\{ t_{AA} (1-n_{i \bar{\sigma}}) (1-n_{j 
\bar{\sigma}}) \right. \nonumber \\ &&+ \left. t_{BB} \ n_{i \bar{\sigma}}n_{j 
\bar{\sigma}} + t_{AB} \left[ n_{i \bar{\sigma}} (1-n_{j \bar{\sigma}}) + 
n_{j \bar{\sigma}} (1-n_{i \bar{\sigma}}) \right] \right\} \nonumber \\ 
&&+ \ U \sum_i 
(n_{i \uparrow} -\frac{1}{2}) (n_{i \downarrow} -\frac{1}{2}) \ .
\label{eq1}
\end{eqnarray}
$H$ contains the most general form of hopping term describing the 
low energy physics of a broad class of system Hamiltonians in which 
four states per effective site are retained. In particular, the 
Hamiltonian $H$ in Eq. (1) has been derived and studied for transition 
metals, organic molecules and compounds \cite{str}, intermediate 
valence systems, cuprates and other superconductors \cite{sev}.
In the continuum limit, the only relevant interactions at 
half-filling are $U$ and $t_{AA}+t_{BB}-2t_{AB}$ \cite{japa}. Therefore, we 
restrict the present study to the electron-hole symmetric case 
($t_{AA}=t_{BB}=1$) which has spin and pseudospin SU(2) symmetries, 
the latter with generators $\eta^+ = \sum_i (-1)^i c^{\dagger}_{i 
\uparrow} c^{\dagger}_{i \downarrow}$, $\eta^- = (\eta^+)^{\dagger}$, and 
$\eta^z = \frac{1}{2} \sum_i(\sum_{\sigma} n_{i \sigma} - 1)$.
The canonical transformation (CT) $\tilde{c}_{i \uparrow}=c_{i \uparrow}$, 
$\tilde{c}_{i \downarrow}=(-1)^i c^{\dagger}_{i \downarrow}$ changes the 
sign of $U$ in $H$, and interchanges the total spin and pseudospin operators 
($\eta^{\alpha} \longleftrightarrow S^{\alpha}$). These symmetry properties 
become crucial in this work. For $t_{AB}$ =0, the 
model has been solved exactly \cite{afg} with the result that the ground 
state (GS) is highly degenerate. For $t_{AB} \neq 0$ the physics of the model 
is still unclear and constitutes our main concern. 

For the present case $\gamma_{c,s}$ are defined as \cite{gag}

\begin{eqnarray}
\gamma_{c,s} &=& i \int_0^{2\pi} d\phi \ \langle g_K(\phi,\pm \phi) | 
\partial_{\phi} g_K(\phi,\pm \phi) \rangle \ , 
\end{eqnarray}
where $|g_K(\phi_{\uparrow},\phi_{\downarrow})\rangle$ is the GS 
in the subspace with total wave vector $K$ and other quantum numbers kept 
fixed, with fluxes $\phi_{\sigma}$ for spin $\sigma$. 
Changes in macroscopic polarization with spin $\sigma$, $P_{\sigma}$, 
are related to the corresponding changes in the Berry phase by: 
$\Delta P_{\uparrow} \pm \Delta P_{\downarrow}= e \Delta 
\gamma_{c,s}/2\pi$ (mod($e$)) \cite{ort,gag,epl}. 
Thus, a phase transition will be detected by a jump in $\pi$ of
$\gamma_c$ ($\gamma_s$) if and only if both thermodynamic phases
differ in $P_{\uparrow} + P_{\downarrow}$ ($P_{\uparrow} - 
P_{\downarrow}$) by $e/2$ (mod($e$)). For example, if one of the 
phases is a CDW with maximum order parameter (CDWM) and the other a 
N\'eel state (N), one is transformed into the other transporting half 
of the charges (those with a given spin) one lattice parameter. In 
addition, as explained below, in the present model topological 
transitions in $\gamma_c$ and $\gamma_s$ indicate the opening of the 
charge and spin gap $\Delta_c$, $\Delta_s$.

We find that the 
minimum of the GS energy as a function of fluxes,  
$E_g(\phi_{\uparrow},\phi_{\downarrow})$, corresponds to the so-called 
closed shell conditions (CSC): if the number of sites (assumed even) 
is $L=2$ (mod(4)), then $K=\phi_{\sigma}=0$, while for $L$ multiple of four 
$K=\phi_{\sigma}=\pi$ (which is equivalent to taking antiperiodic boundary 
conditions and $\bar{K}=0$ in a system without fluxes). 


\begin{figure}
\narrowtext
\epsfxsize=3.5truein
\vbox{\hskip 0.05truein \epsffile{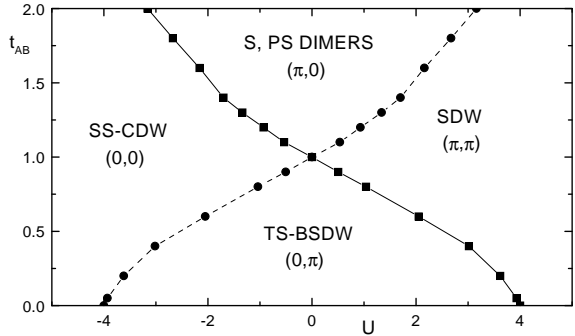}}
\medskip
\vspace*{-0.8cm}
\caption{Quantum phase diagram of the correlated hopping Hubbard chain.
The vector Berry phase $\vec{\gamma}=(\gamma_c,\gamma_s)$ and the nature 
of each stable phase is indicated: Luther-Emery liquid with equally 
decaying singlet superconductor and CDW correlations (SS-CDW), 
Tomonaga-Luttinger liquid with triplet superconducting and 
bond SDW correlations dominating at large distances (TS-BSDW), SDW 
insulator (SDW), and spin and pseudospin dimerization (S,PS).}
\label{fig1}
\end{figure}

For any finite system and fixed $t_{AB}$, varying $U$, 
two topological transitions occur in the 
model, corresponding to a jump in either 
$\gamma_c$ or $\gamma_s$. We have determined those transitions in rings of 
length $L=6,8,10,12$ using the Lanczos method. The results extrapolated with 
a cubic polynomial in $1/L$ are represented in Fig.~\ref{fig1}. In contrast 
to other physical quantities which show large finite-size effects, 
particularly near $t_{AB}=0$ \cite{mit}, the topological transitions converge 
rapidly to the thermodynamic limit (for example, for $t_{AB}=0.05$, 
$\gamma_c$ jumps at 
$U$=3.451, 3.681, 3.788, and 3.846 for $L=6-12$, and the extrapolated value 
is $U$=3.932). The numerical convergence becomes problematic for smaller 
$t_{AB}$ values. At $t_{AB}=0$ the transition points are determined from the 
exact solution \cite{afg} as those values of $U$ where  
$\Delta_c$ and $\Delta_s$ open. Those critical 
values are $U_{c,s}=\pm 4$ and match smoothly with the rest of the curves in 
Fig.~\ref{fig1}. It is easy to see that under CT the geometrical phases 
transform as $\gamma_c \longleftrightarrow \gamma_s + \pi$ \cite{gag}. Thus, 
as seen in Fig.~\ref{fig1}, a jump in $\gamma_c$ at $U_c$ (full line) implies a 
jump in $\gamma_s$ at $-U_c$ (dashed line), and vice versa. 

In the case where all particles are localized one can easily determine 
the value of $\vec{\gamma}$ as 
$\gamma_{c,s} = \textrm{Im} \ln z_L^{c,s}$, where 
\begin{equation}
z_L^{c,s} = \langle g | e^{i \frac{2\pi}{L} \sum_j j (n_{j \uparrow} \pm 
n_{j \downarrow})} | g \rangle \ ,
\end{equation}
was recently used to study quantum localization \cite{zres,znos}. 
In the thermodynamic limit $z_L^c$ vanishes for a conductor while $|z_L^c| 
\rightarrow 1$ for an insulator. Clearly, 
$\vec{\gamma}(\rm CDWM)=(0,0)$, while $\vec{\gamma}(\rm N)=(\pi,\pi)$, and by
continuity $\vec{\gamma}(\rm CDW)=(0,0)$, $\vec{\gamma}(\rm SDW)=(\pi,\pi)$.
On the other hand, it is not easy to predict 
the values of $\vec{\gamma}$ in conducting phases. However, for $U=0$, the 
model is invariant under CT. Therefore $\vec{\gamma}=(\pi,0)$, or 
$\vec{\gamma}=(0,\pi)$, indicating a topological difference with the above 
mentioned CDW and SDW states. 

\begin{figure}
\narrowtext
\epsfxsize=3.5truein
\vbox{\hskip 0.05truein \epsffile{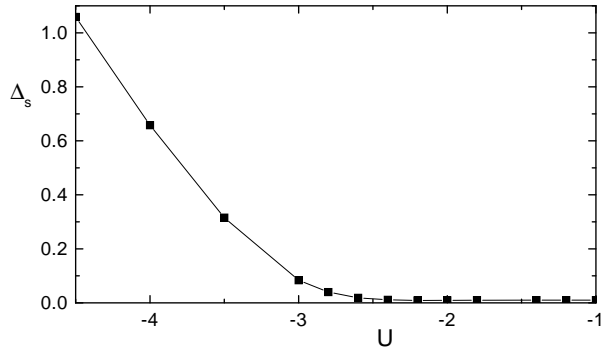}}
\medskip
\vspace*{-0.5cm}
\caption{Spin gap $\Delta_s$ as a function of $U$ for $t_{AB}=0.6$, obtained 
with DMRG.}
\label{fig2}
\end{figure}

$\Delta_c$ ($\Delta_s$) vanishes at the left (right) of the full (dashed) 
line in Fig.~\ref{fig1}, and is different from zero at the right (left). It 
has been shown,
that in spin SU(2) symmetric systems the opening of 
$\Delta_s$ can be detected as a level crossing of triplet and singlet states 
for boundary conditions opposite to the CSC ones (periodic if the number of 
particles with a given spin $N_{\sigma}$ is even, and antiperiodic if 
$N_{\sigma}$ is odd) \cite{nak}.
It is precisely this crossing which causes
the jump in $\gamma_s$ \cite{epl}. 
A direct evaluation of $\Delta_s$ has large finite-size effects, and an 
accurate calculation requires use of the Density Matrix Renormalization 
Group (DMRG) method \cite{white}. Fig. \ref{fig2} displays $\Delta_s$ for $t_{AB}=0.6$ 
extrapolated from calculations of open chains with length $L \leq 40$. For 
$-2 < U < -1$, $\Delta_s$ vanishes within numerical accuracy ($\sim 0.01$) 
while for smaller values of $U$ it increases rapidly. These results are 
consistent with $\Delta_s$ opening near $U_s \approx -2.1$ with a 
singular growth (exponentially small as in the $t-J$ model \cite{nak}). 
The result derived from the topological transition is $U_s=-2.051$.

\begin{figure}
\narrowtext
\epsfxsize=3.0truein
\vbox{\hskip 0.05truein \epsffile{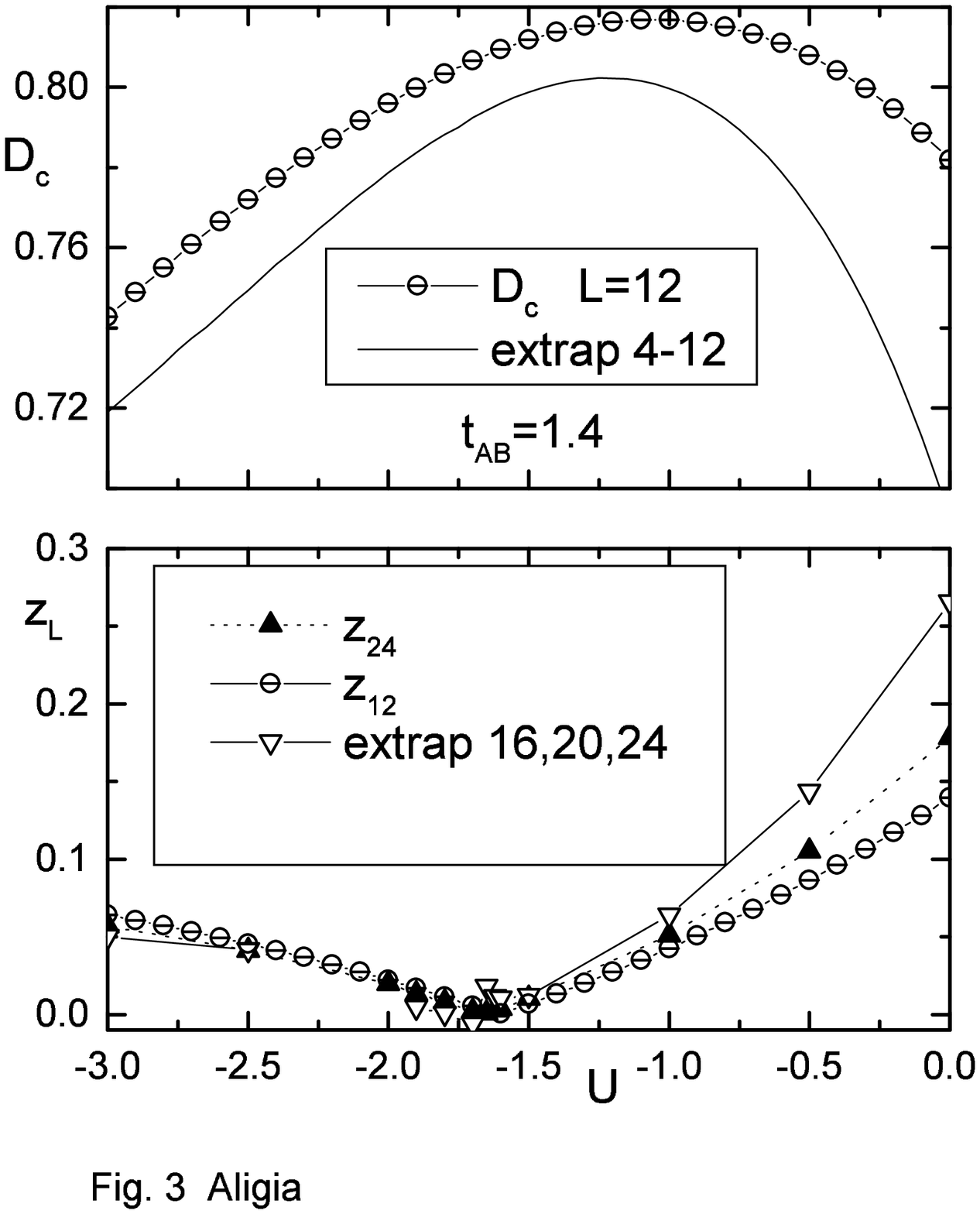}}
\medskip 
\vspace*{-2cm}
\caption{Drude weight $D_c$ and $z_L^c$ for $t_{AB}=1.4$ 
as a function of $U$ for $L=12$ (open circles) and $L=24$ (solid triangles).
The solid lines are polynomial extrapolations in $1/L$.}
\label{fig3}
\end{figure}

The symmetry transformation CT implies that if $\Delta_s$ opens at $U_s$, 
a pseudospin gap $\Delta_{\eta}$ opens at $-U_s$. Pseudospin excitations 
form a subset of all charge excitations. In our case the charge 
velocity $v_c$ and $\Delta_c$ (computed as usual in finite systems $L \leq 12$
\cite{nak}) coincide with their pseudospin counterparts $v_{\eta}$ and 
$\Delta_{\eta}$ (computed from $v_s$ and $\Delta_s$ for opposite $U$) 
\cite{note}. This is consistent with the exact solution for $t_{AB}=0$ 
where the charge excitations of lowest energy are pseudospin ones \cite{afg}.
In addition, the charge-charge and spin-spin correlation functions are 
interchanged as $U$ changes sign (see below). 
The opening of $\Delta_c$ where $\gamma_c$ jumps 
from 0 to $\pi$ is also consistent with calculations of $z_L^c$, $D_c$, 
superconducting correlation functions, $K_{\rho}$ and central charge $c$.
For $t_{AB} > 1$, $\Delta_c$ opens more slowly and 
the detection of the transition
becomes more difficult. However, as shown in Fig.~\ref{fig3}, 
$D_c$ and $z_L^c$ display a 
similar behavior near the jump in $\gamma_c$ ($U_c$=-1.702) as in the 
quarter-filled infinite $U$ extended Hubbard model as a function of the
nearest-neighbor repulsion $V$ \cite{znos}, where a metal-insulator 
transition takes place at $V = 2 t$. At large $|U|$ the only relevant energy 
scale is $4 t_{AB}^2/|U|$ and therefore, $D_c$ increases with $U$ for large 
negative $U$. Near $U_c$ there is a drastic change of behavior of $D_c$ 
and $z_L$ vs $U$, and for $U > U_c$, the extrapolated values suggest a tendency
to reach the insulating values ($D_c$=0, $z_L^c$=1), 
in the thermodynamic limit.
$z_L^c$ always decreases (increases) with $L$ at the left (right) of $U_c$.
This is the first accurate (DMRG) calculation of $z_L^c$ in a system of
more than 16 sites.

In order to further characterize each thermodynamic phase we use 
symmetry arguments and the following correlation functions (CF) (see Fig. 4):
\begin{eqnarray}
\chi_{t(s)}(d) &=& \frac{1}{2} \langle (c^{\dagger}_{0 \uparrow} 
c^{\dagger}_{1 \downarrow} \pm c^{\dagger}_{0 \downarrow} 
c^{\dagger}_{1 \uparrow})(c^{\;}_{d+1 \downarrow} 
c^{\;}_{d \uparrow} \pm c^{\;}_{d+1 \uparrow} 
c^{\;}_{d \downarrow}) \rangle \nonumber \\
\chi_{os}(d) &=& \langle c^{\dagger}_{0 \uparrow} 
c^{\dagger}_{0 \downarrow} c^{\;}_{d \downarrow} c^{\;}_{d \uparrow} \rangle 
\nonumber \\
\chi_{c}(d) &=& -\frac{1}{2} \langle (n_{0 \uparrow} + n_{0 \downarrow} 
- 1)(n_{d \uparrow} + n_{d \downarrow} - 1) \rangle \nonumber \\
\chi_{bsdw}(d) &=& \frac{1}{2} \langle (c^{\dagger}_{0 \uparrow} 
c^{\;}_{1 \downarrow} + c^{\dagger}_{1 \uparrow} 
c^{\;}_{0 \downarrow})(c^{\dagger}_{d+1 \downarrow} 
c^{\;}_{d \uparrow} + c^{\dagger}_{d \downarrow} 
c^{\;}_{d+1 \uparrow}) \rangle \nonumber \\
\chi_{sz}(d) &=& -2 \langle S_0^z S_d^z \rangle \ ,
\label{eq4}
\end{eqnarray}  
computed with DMRG and CSC \cite{note}. 
The CF are defined such that, in 
the non-interacting case, $\lim_{L \rightarrow \infty} \chi(d,L) = 
1/(\pi d)^2$, with $d$ an odd number. $\chi_{bsdw}(d)$ corresponds to 
correlations between spins located in bonds \cite{japa}.
The GS for CSC is always a spin and pseudospin singlet. Invariance 
under pseudospin rotations implies that $\chi_{os}(d) = (-1)^{d+1} \chi_{c}(d)$, 
i.e. on-site singlet superconducting and charge CF 
have the same distance dependence. This, in turn, implies $K_{\rho}=1$ in the 
conducting phases \cite{voit}, meaning that 
the large distance behavior is dominated by logarithmic corrections.
Also, using CT and spin rotations 
one obtains $\chi_{c}(d,U)=\chi_{sz}(d,-U)$ and 
$\chi_{t}(d,U)=\chi_{t}(d,-U)=(-1)^d \chi_{bsdw}(d,U)$.

\begin{figure}
\narrowtext
\epsfxsize=3.0truein
\vbox{\hskip 0.05truein \epsffile{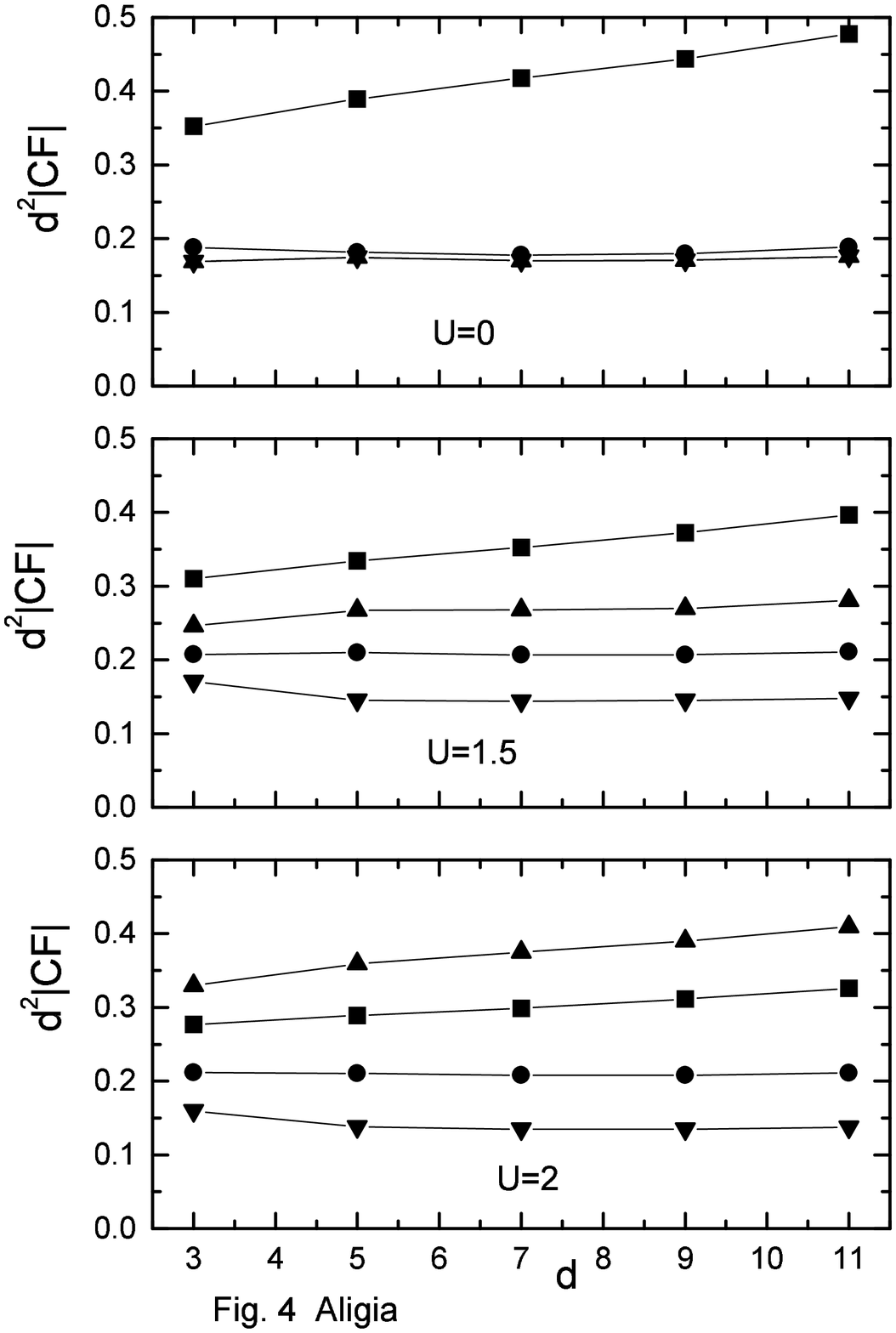}}
\medskip
\caption{Correlation functions (CF) times the square of the distance $d$ 
as a function of $d$ for $t_{AB}=0.6$: nearest-neighbor triplet 
$|\chi_{t}(d)|=|\chi_{bsdw}(d)|$
(squares) and singlet $\chi_{s}(d)$ (circles) pair CF, on-site pair and 
charge CF $|\chi_{os}(d)|=|\chi_{c}(d)|$  (downward triangles), 
and spin CF $\chi_{sz}(d)$ (upward triangles). 
See Eq.~(\ref{eq4}).}
\label{fig4}
\end{figure}

We restrict our study to odd values 
of $d$ for which the oscillatory factors of the CF in the continuum limit 
are maximum \cite{voit}. 
In addition we took $L=2d$, motivated by results on the  
Heisenberg model \cite{kar1}, showing that $\lim_{L \rightarrow \infty} 
\chi_{sz}(d,L)= C \chi_{sz}(d,2d)$, where $C$ is a constant 
independent of $d$. This also true, if $d$ is odd, for all CF of our model in the 
non-interacting case, with $C=(\pi/2)^2$. 
The above mentioned symmetry relations and analytical 
results in the non-interacting limit 
allowed us to check the accuracy of the DMRG results. A summary for each 
non-equivalent topological region is described below:

a) $\underline{\vec{\gamma}=(\pi,0), \Delta_c \neq 0, \Delta_s \neq 0.}$ 
All CF decay exponentially. To understand the nature of this insulating GS 
let us analyze the limit $t_{AB} \rightarrow \infty$, where effectively 
there is a sequence of spin and pseudospin dimers. Each dimer is the
ground state of the model for two sites and two particles. 
Including the dimer-dimer 
interaction in second-order perturbation theory the resulting GS 
energy per site is $e \approx -1.0625 t_{AB}$ 
($e_{\rm DMRG} = -1.0808 t_{AB}$). 
The energy difference between the two lowest singlet states 
(with $\bar{K}=0$ and $\bar{K}=\pi$) decays 
nearly exponentially with $L$, indicating a breaking 
of the translational symmetry in the thermodynamic limit.

b) $\underline{\vec{\gamma}=(\pi,\pi), \Delta_c \neq 0, \Delta_s = 0.}$
As in the ordinary positive $U$ Hubbard model, the GS is an insulating SDW. 

c) $\underline{\vec{\gamma}=(0,0), \Delta_c = 0, \Delta_s \neq 0.}$
The system forms a Luther-Emery liquid with $\chi_s$ and $\chi_c$ CF 
decaying as $1/d$ (neglecting logarithmic corrections.) 

d) $\underline{\vec{\gamma}=(0,\pi), \Delta_c = 0, \Delta_s = 0.}$
$\chi_{os}$ and $\chi_{s}$ display the 
same long distance behavior (Fig.~\ref{fig4}). Because of 
symmetry arguments, for $U=0$, all CF except $\chi_t=\chi_{bsdw}$ 
show the same $1/d^2$ 
decay (apparently without logarithmic corrections). Renormalization group 
arguments \cite{voit} show that when $\Delta_s=0$, $\chi_t$ should decay 
more slowly than $\chi_s$, and therefore dominate 
for small $|U|$. As $U$ increases, $\chi_t$ decays more rapidly, while the 
opposite happens with $\chi_{sz}$. Near the opening of $\Delta_c$ (at 
$U_c \approx 2.05$ for $t_{AB} = 0.6$), both CF seem to decay in a similar 
fashion. 
               
In conclusion, we have constructed the quantum phase diagram of Eq.~(\ref{eq1})
from topological considerations. Each thermodynamic phase is associated to a 
topological vector, and changes in that quantity signal the 
transition point. A key finding is the identification of a triplet 
superconducting phase degenerate with a bond-located SDW for 
$t_{AB} < 1$ and small $|U|$. Taking into account the slight 
dimerization in (TMTSF)$_2$X compounds, an effective Hamiltonian 
similar to Eq. (\ref{eq1}) can be realized, where only four low-energy 
states per unit cell are kept. Assuming that additional interactions 
stabilize the TS with respect to the BSDW it is remarkable that, 
for $U > 0$, as $t_{AB}$ is decreased a similar sequence of quantum 
phase transitions takes place as pressure is applied \cite{jer}. This 
general approach can be extended to spatial dimensions higher than one 
\cite{ort}, and applied to more general models which do not 
necessarily have SU(2) pseudospin symmetry \cite{gag}.

One of us (A.A.A.) thanks J. Voit and L. Arrachea for helpful 
discussions. We acknowledge computer time at the Max-Planck 
Institute f\"ur Physik Komplexer Systeme. A.A.A., K.H. and C.D.B. are 
supported by CONICET, Argentina, while G.O. is supported by the US 
Department of Energy. 
 


\begin{references}
\bibitem{note}
For boundary conditions other than open or CSC the DMRG is not accurate 
enough for our model. Thus, calculations of $\vec{\gamma}$, $D_c$, $v_{c,s}$, 
and $K_{\rho}$ were done using the Lanczos method with $L \leq 12$.
\end{references}
\end{document}